\documentclass[%
 reprint,
 amsmath,amssymb,
 aps,
]{revtex4-2}

\usepackage{graphicx}
\usepackage{dcolumn}
\usepackage{bm}
\usepackage{array} 

\begin{document}

\preprint{APS/123-QED}

\title{Spontaneous Symmetry Breaking and Collective Higgs–Goldstone Dynamics in Solid-State Phononic Frequency Combs}

\author{Murtaza Rangwala$^{1}$ and Adarsh Ganesan$^{1,2}$}
\affiliation{$^{1}$Department of Electrical and Electronics Engineering, Birla Institute of Technology and Science, Pilani - Dubai Campus, Dubai International Academic City, Dubai, UAE 345055}
\email{Email: adarsh@dubai.bits-pilani.ac.in}
\affiliation{$^{2}$Department of Mechanical Engineering, Birla Institute of Technology and Science, Pilani - Pilani Campus, Vidya Vihar, Pilani, India 333031}

\date{\today}

\begin{abstract}
We investigate the generation of phononic frequency combs arising from nonlinear coupling between Higgs-like and Goldstone-like phonon modes in hexagonal InMnO$_3$. The Higgs-like mode, an infrared-active optical phonon, is resonantly driven by a short, high-electric field terahertz pulse, while the optically inactive Goldstone-like mode is indirectly excited through intrinsic nonlinear mode coupling. Using a nonlinear phononics model, we numerically solve the coupled equations of motion governing the lattice dynamics and analyze the resulting time- and frequency-domain responses. By systematically varying key drive and material parameters—including electric field amplitude, pulse width, driving frequency, and mode damping—we identify the conditions under which stable phononic frequency combs emerge. Our results reveal clear threshold behaviors for comb formation, tunability of comb spacing and spectral bandwidth through external control parameters, and a breakdown of coherent comb structure at high drive strengths or weak damping. These findings demonstrate how nonlinear Higgs--Goldstone interactions enable controllable phononic frequency comb generation and provide insight into ultrafast lattice dynamics in symmetry-broken materials.
\end{abstract}

\maketitle

\section{Introduction} 
Phononic frequency combs (PFCs) are spectral structures composed of evenly spaced, phase-coherent frequency lines generated through nonlinear interactions among mechanical vibrational modes \cite{cao2014phononic,ganesan2017phononic,ganesan2018excitation,ganesan2018phononic,goryachev2020generation,de2023mechanical,sun2023generation,lei2025nonlinear}. Much like optical frequency combs, which revolutionized precision metrology and spectroscopy, PFCs offer a pathway to engineer and interrogate mechanical dynamics with unprecedented spectral resolution. Their formation originates from nonlinear mode coupling, parametric resonance, or higher-order interactions, making them sensitive probes of intrinsic material properties.

Beyond the nonlinear dynamics that generate PFCs, recent advances in solid-state phonon engineering have highlighted the broader potential of manipulating lattice vibrations across a wide frequency spectrum. For example, atomically thin van der Waals heterostructures have been used to generate and detect terahertz phonons, achieving frequency-comb-like multimode Fabry--P\'erot spectra with more than 20 overtone peaks \cite{yoon2024terahertz}. In such architectures, few-layer graphene serves as an ultrabroadband phonon transducer, converting femtosecond near-infrared pulses into high-frequency acoustic waves, while monolayers of WSe$_2$ act as sensitive phonon detectors through strong exciton--phonon coupling. These advances enable terahertz phononic spectroscopy, high-Q THz phononic cavities, and direct extraction of heterointerface force constants.

Chen et al.~(2024) \cite{chen2025spontaneously} reported the spontaneous formation of long-lived PFCs in van der Waals materials CrGeTe$_3$ and CrSiTe$_3$, observed via high-resolution Raman scattering. In these compounds, a localized Ag5 optical phonon splits into a series of equidistant, sharp peaks that persist for hundreds of oscillations and survive up to 100 K. The responsible phonon branch exhibits a nearly flat dispersion, corresponding to localized oscillations of Ge$_2$Te$_6$ or Si$_2$Te$_6$ clusters. A simple cubic anharmonicity in a single-mode oscillator model was sufficient to replicate the observed comb structure, illustrating that flat, localized modes can act as nearly independent quantum oscillators.

While prior work \cite{yoon2024terahertz,chen2025spontaneously} has largely explored PFCs in van der Waals materials, the present study investigates PFC generation through nonlinear interactions among phonon modes in solid-state crystals \cite{forst2011nonlinear,subedi2014theory}. Over the past 15 years, nonlinear phononics has enabled selective vibrational excitation by deforming the lattice along specific normal mode coordinates \cite{mankowsky2016non,subedi2017midinfrared,gomez2022light,tang2023light,shin2024light}, including the demonstration of light-induced superconductivity \cite{fausti2011light,mitrano2016possible,cantaluppi2018pressure}. Using multidimensional THz spectroscopy, researchers have directly imaged the amplitude--phase dynamics of light-enhanced superconducting states, revealing collective oscillations of the order parameter, including mixed phase--amplitude modes and higher-order sidebands. These studies highlight the importance of precision tools that access discrete, coherence-preserving frequency structures. PFCs provide a naturally suited platform for such investigations, as discrete, evenly spaced comb lines generated by nonlinear phonon interactions can probe symmetry-broken lattices and selectively drive or read out phonon modes. Mode-selective optical control allows coherent excitation of one or a few low-energy degrees of freedom, in contrast to incoherent near-visible excitation, which is less selective and induces heating.

Achieving PFCs in solid-state systems, however, remains challenging. Many materials possess weak intrinsic nonlinearities, strong damping, or poorly tuned modal separations that inhibit parametric amplification. Systems with engineered symmetry breaking, defect-induced confinement, or tunable elastic landscapes offer more favorable conditions but still require careful adjustment of drive amplitude, detuning, coupling strengths, and modal frequency spacing. A central objective of this work is to map these parameter regimes and identify conditions under which solid-state systems naturally support PFC formation.

This challenge connects to a deeper conceptual framework: the role of order parameters in describing collective degrees of freedom. Order parameters encode amplitudes and phases that can be driven far from equilibrium by external stimuli, such as laser pulses or mechanical drives. Their excitations include Higgs and Goldstone modes, corresponding to amplitude and phase oscillations of a symmetry-breaking order parameter. Initially developed in particle physics \cite{higgs1964broken}, these concepts have been observed in superfluids \cite{behrle2018higgs}, supersolids \cite{guo2019low}, superconductors \cite{cea2015nonrelativistic}, charge-density-wave materials \cite{cea2014nature}, antiferromagnets \cite{jain2017higgs}, and excitonic insulators \cite{golevz2020nonlinear}. Recent theory suggests that optical and acoustic vibrational modes of solids themselves can behave as Higgs-like and Goldstone-like excitations of the crystal lattice, particularly in materials with Mexican-hat--like energy landscapes \cite{juraschek2020parametric, vallone2020higgs}.

In this context, hexagonal InMnO$_3$ is a compelling platform. Its unique structural properties realize a two-component order parameter that couples MnO$_5$ tilting and In buckling, forming a potential energy landscape that can be approximated by a ``Mexican hat''. Upon cooling, the high-symmetry, nonpolar state transitions to a set of low-energy minima, corresponding to discrete—but closely spaced—ground states with a fixed amplitude $|\eta|$ and a selected phase $\Phi_0$. Fluctuations around these states give rise to collective excitations: the high-energy Higgs mode corresponds to amplitude oscillations of the coupled distortions, while the low-energy Goldstone mode corresponds to phase oscillations along the energy minima. The trilinear coupling in the P6$_3$cm crystal structure locks these distortions together, forming a complex order parameter that underpins the observed symmetry-breaking behavior. In this work, we explore how external drive parameters—including pulse width ($\tau$), electric field intensity ($E_0$), carrier frequency ($f_D$), and damping coefficient ($\gamma$)—can excite these modes and induce PFCs, providing insight into controlled manipulation of coupled lattice dynamics. To investigate this, we study the nonlinear coupled differential equations governing interacting phonon modes to generate time--frequency spectra, parametric sweeps, and contour maps that reveal how tuning physical parameters controls comb onset and spacing.

\section{Model and Numerical Setup}

The present study is based on the nonlinear phononics framework developed for hexagonal InMnO$_3$, where Higgs-like and Goldstone-like phonon modes emerge from a symmetry-broken lattice potential. In this system, the Higgs-like mode is an infrared-active optical phonon that can be directly excited by an external electric field, while the Goldstone-like mode is optically inactive but nonlinearly coupled to the Higgs mode. As demonstrated in prior work \cite{juraschek2020parametric}, a short-lived, high-field terahertz pulse can resonantly drive the Higgs mode, which in turn transfers energy to the Goldstone-like mode through intrinsic nonlinear coupling between the two lattice distortions.

In this work, we extend this framework to explore the generation of phononic frequency combs arising from the nonlinear interaction between these coupled phonon modes.

The coupled lattice dynamics of the Higgs-like and Goldstone-like phonon modes in hexagonal InMnO$_3$ are described by two nonlinear second-order differential equations for the mode coordinates $x_1(t)$ and $x_2(t)$. These equations capture harmonic restoring forces, dissipation, nonlinear mode coupling, and external driving by a terahertz electric field. Explicitly, the equations of motion are given by
\begin{align}
\ddot{x}_1(t) &=
-2\frac{1}{Q_1}\dot{x}_1(t)
- \omega_1^2 x_1(t)
- \alpha_{22} x_2^2(t) \nonumber \\
&\quad
+ Z_h E_0
\exp\!\left[-\frac{4\ln 2}{\tau^2} t^2 \right]
\cos(\omega_D t),
\label{eq:higgs} \\
\ddot{x}_2(t) &=
-2\frac{1}{Q_2}\dot{x}_2(t)
- \omega_2^2 x_2(t)
- \alpha_{12} x_1(t) x_2(t).
\label{eq:goldstone}
\end{align}

Here, $x_1(t)$ represents the Higgs-like phonon coordinate and $x_2(t)$ the Goldstone-like phonon coordinate. The parameters $\omega_1$ and $\omega_2$ denote the natural angular frequencies of the Higgs and Goldstone modes, respectively, while $Q_1$ and $Q_2$ are their quality factors, accounting for intrinsic damping. The coefficients $\alpha_{22}$ and $\alpha_{12}$ describe the nonlinear coupling between the two modes, enabling energy transfer from the driven Higgs mode to the optically inactive Goldstone mode. The values of modal frequencies and nonlinear coupling coefficients associated with InMnO\textsubscript{3} are obtained from \cite{juraschek2020parametric}.

The Higgs-like mode is driven by a linearly polarized terahertz electric field with peak amplitude $E_0$, carrier angular frequency $\omega_D$, and a Gaussian temporal envelope with full width at half maximum (FWHM) $\tau$. The coupling constant $Z_h$ represents the effective mode charge associated with the infrared-active Higgs phonon. The Goldstone-like mode is not directly driven but becomes excited through its nonlinear interaction with the Higgs mode.

To investigate the conditions under which phononic frequency combs emerge, we systematically vary key drive and material parameters, including the pulse width, electric field amplitude, driving frequency, and damping coefficients of both phonon modes. For each parameter set, the coupled equations of motion are solved numerically using the \texttt{SciPy} library in Python. Specifically, we employ the implicit Radau method, which is well suited for stiff nonlinear systems and allows stable time evolution over the relevant dynamical regimes. The equations are integrated over carefully chosen temporal windows to ensure both transient and steady-state dynamics are captured.

The parameter ranges explored in this study are listed in TABLE I.

\begin{table}[ht]
\centering
\caption{Parameter ranges explored in the numerical simulations for contour-plot generation.}
\label{tab:parameter_ranges}
\begin{ruledtabular}
\begin{tabular}{lll}
\textbf{Parameter} & \textbf{Symbol} & \textbf{Range} \\
\hline\\
Quality factor & $Q_{1,2}$ & $[10,\,2000]$ \\\\
Peak electric field & $E_0 (MV/m)$ & $\left[26.9507,\,46.6454\right]$ \\\\
Pulse width (FWHM) & $\tau (ps)$ & $[0.5,\,2.0]$\\\\
Driving frequency & $f_D (THz)$ &
$\left[3.4,4.6\right]$ \\\\
\end{tabular}
\end{ruledtabular}
\end{table}





For each parameter set, the time evolution of $x_1(t)$ and $x_2(t)$ is computed, yielding displacement--time traces for both phonon modes. These time-domain signals are then Fourier transformed to extract the corresponding frequency-domain spectra. The resulting amplitude-versus-frequency distributions reveal the emergence of evenly spaced spectral lines characteristic of phononic frequency combs. By analyzing these spectra across the explored parameter space, we determine how variations in drive strength, pulse duration, driving frequency, and damping influence the onset of comb formation, the spacing between comb lines, and the total number of generated frequency components.

Finally, to visualize these dependencies, we construct contour maps that directly correlate control parameters with comb characteristics such as line spacing and spectral density. These maps provide a clear overview of the parameter regimes that favor robust phononic frequency comb generation and highlight the central role of nonlinear Higgs--Goldstone coupling in shaping the observed spectral structures.

\section{Results and Discussion}

 \begin{figure*}
    \centering
    \includegraphics[width=0.7\linewidth]{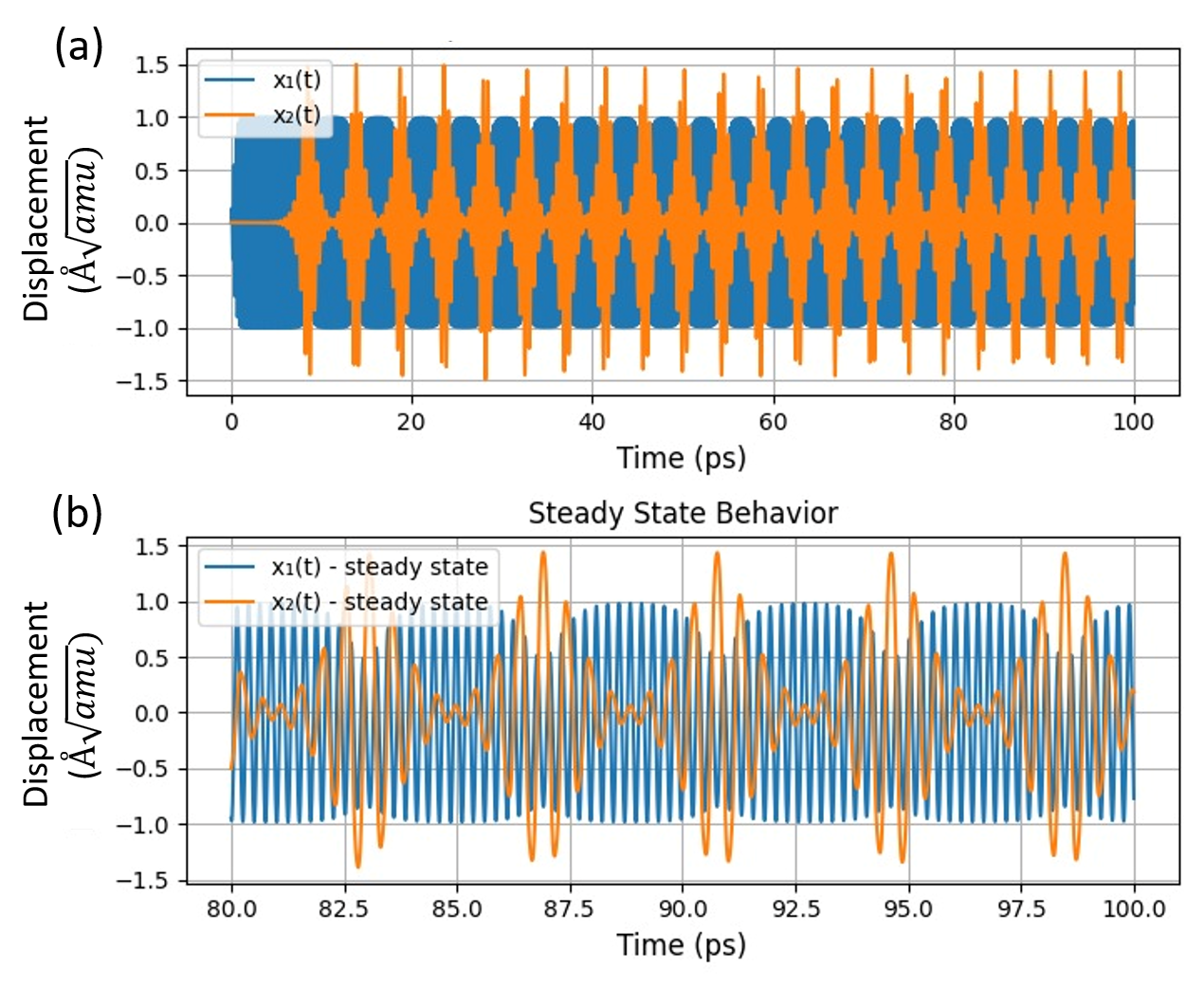}
    \caption{Time-domain dynamics of the coupled Higgs-like and Goldstone-like phonon modes in hexagonal InMnO$_3$.
    The figure shows the displacement $x_1(t)$ of the infrared-active Higgs mode driven by a terahertz pulse, $x_2(t)$ of the Goldstone-like mode excited through nonlinear coupling.(a) shows the complete time domain response spanning 100ps, (b) shows the steady state response, the last 20\% of the time doamin response, refer TABLE II for the values used.}
    \label{fig1}
\end{figure*}
\subsection{Time-Domain Dynamics and Steady-State Behavior}

Fig.~1 shows the time-domain dynamics of the coupled Higgs-like and Goldstone-like phonon modes obtained by numerically solving the nonlinear equations of motion. The system is initialized with the conditions,
\[
[x_1(0),\,\dot{x}_1(0),\,x_2(0),\,\dot{x}_2(0)] = [0,\,0,\,10^{-5},\,0]\mathrm{~\AA\sqrt{amu}},
\]
and the equations are integrated over a total time window of 100~ps.

The upper panel (a) displays the full displacement--time evolution of both phonon coordinates. Upon application of the terahertz electric pulse, the optically active Higgs-like mode is directly excited and reaches a maximum displacement of approximately $1~\mathrm{~\AA\sqrt{amu}}$. After the pulse has passed, the Higgs mode continues to oscillate at its natural frequency. Owing to the intrinsic nonlinear coupling between the two modes, energy is progressively transferred from the Higgs mode to the optically inactive Goldstone-like mode, which subsequently begins to oscillate with increasing amplitude.

A clear signature of this energy exchange is observed in the anti-correlated oscillation amplitudes of the two modes: peaks in the Goldstone-mode displacement coincide with dips in the Higgs-mode oscillation amplitude. This behavior reflects periodic energy transfer between the coupled degrees of freedom and constitutes a hallmark of nonlinear mode coupling.

Such energy exchange has close analogs in the mass-spring model. For instance, a mass attached to a spring can exhibit coupled translational and torsional motion, where energy oscillates between vertical displacement and twisting modes. In an analogous manner, the Higgs and Goldstone phonons behave as coupled oscillators that continuously exchange energy through nonlinear interactions.

The lower panel (b) of Fig.~1 shows the steady-state dynamics, obtained by isolating the final 20~ps of the numerical solution. This time window excludes transient effects associated with pulse excitation and highlights the long-lived oscillatory behavior that underlies the formation of well-defined spectral features in the frequency domain.

 \begin{figure*}
    \centering
    \includegraphics[width=0.85\linewidth]{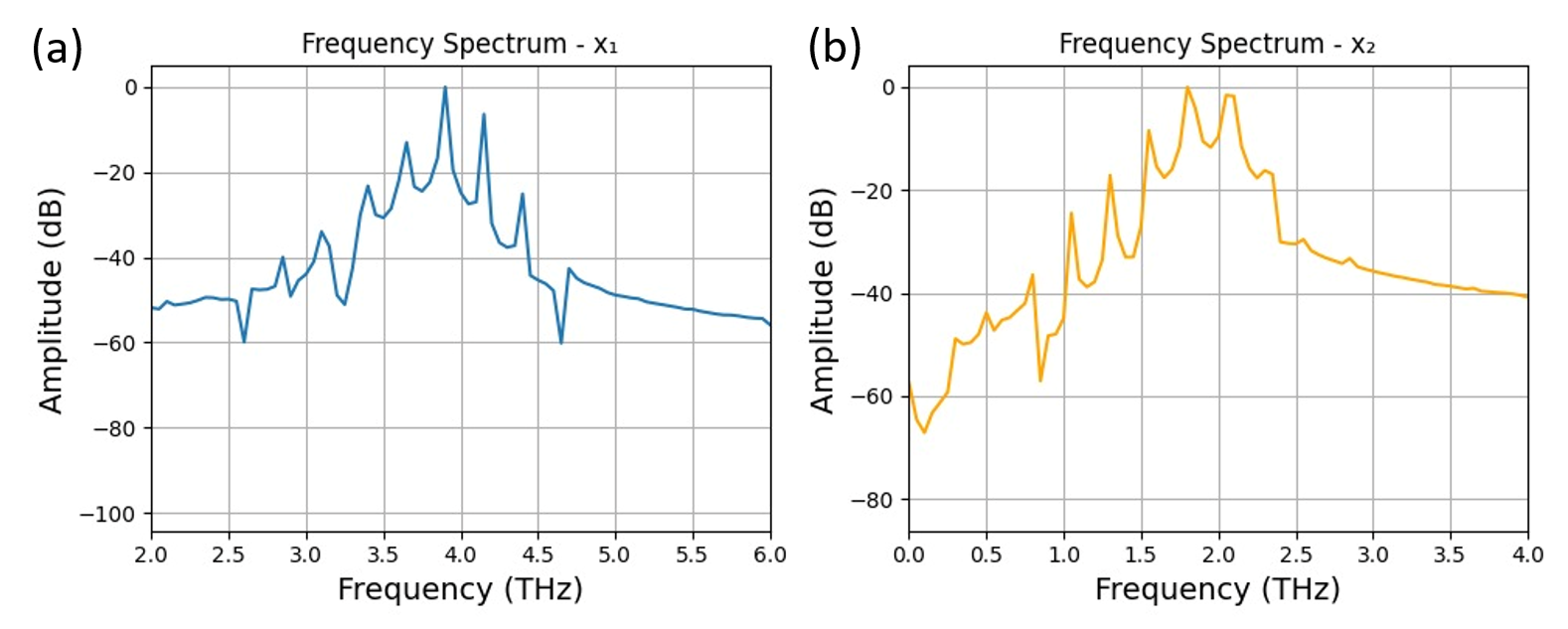}
    \caption{Frequency-domain dynamics of the coupled Higgs-like and Goldstone-like phonon modes. (a) shows the $x_1(t)$ mode and (b) shows the $x_2(t)$ mode. Refer TABLE II for the values used. }
    \label{fig1}
\end{figure*}
\subsection{Frequency-Domain Response and Phononic Frequency Combs}

Fig.~2 shows the frequency-domain spectra of the Higgs-like ($x_1$) and Goldstone-like ($x_2$) phonon modes. These amplitude--frequency plots are obtained by computing the Fast Fourier transform (FFT) of the steady-state portion of the displacement--time signals shown in Fig.~1. Only the final 20~ps of the numerical time evolution are used in order to suppress transient effects associated with the initial pulse excitation.

Both spectra exhibit a series of sharp peaks that are evenly spaced in frequency. These regularly spaced spectral lines constitute phononic frequency combs and arise from the nonlinear coupling between the Higgs and Goldstone modes. The presence of comb structures in both phonon coordinates indicates that the nonlinear interaction not only governs energy exchange in the time domain, but also generates a highly structured and coherent response in the frequency domain.

The FFT amplitudes are expressed in decibel (dB) units, while the frequency axis is given in terahertz (THz). For the Higgs-like mode $x_1$, the dominant spectral peak appears close to its bare resonant frequency of 4~THz. A slight red shift is observed, with the maximum amplitude occurring at a frequency marginally lower than 4~THz. This shift originates from nonlinear effects associated with large-amplitude oscillations and strong mode coupling induced by the intense terahertz driving field. Similar comb features are observed in the Goldstone-like mode $x_2$, centered around its characteristic frequency near 2.4~THz.

The parameters used to generate the time-domain dynamics in Fig.~1 and the corresponding frequency spectra in Fig.~2 are listed in TABLE II.

\begin{table}[ht]
\centering
\caption{Default parameter values used to generate the time-domain dynamics in Fig.~1 and the frequency spectra in Fig.~2.}
\label{tab:default_parameters}
\begin{ruledtabular}
\begin{tabular}{lll}
\textbf{Parameter} & \textbf{Symbol} & \textbf{Value} \\
\hline\\
Higgs-mode frequency & $\omega_1/2\pi$ & $4.0~\mathrm{THz}$ \\\\
Goldstone-mode frequency & $\omega_2/2\pi$ & $2.4~\mathrm{THz}$ \\\\
Nonlinear coupling\\ coefficient & $\alpha_{22}$ & $88.766~\mathrm{~\AA ^{\text{-1}}\sqrt{amu}^{\text{-1}}ps^{\text{-2}}}$\\\\
Cross-coupling\\coefficient & $\alpha_{12}$ & $2\alpha_{22}$ \\\\
Mode effective charge & $Z_h$ & $2.7e $ \\\\
Damping coefficient & $\gamma_{\text{default}}$ & $1.0 \times 10^{-4}~\mathrm{s^{\text{-1}}}$ \\\\
Quality factor & $Q_{\text{default}}$ & $1/\gamma_{\text{default}}$ \\\\
Electric field amplitude & $E_{0,default}$ & $36.2796~\mathrm{MV/m}$ \\\\
Pulse width & $\tau_{\text{default}}$ & $1.0~\mathrm{ps} $ \\\\
Driving frequency & $\omega_{D,\text{default}}$ & $\omega_1$ \\\\
\end{tabular}
\end{ruledtabular}
\end{table}


These values are kept fixed unless otherwise stated when exploring the broader parameter space in subsequent contour-plot analyses. In the construction of the contour plots, only a single control parameter is varied at a time, while all remaining parameters are held fixed at their respective default values listed in TABLE II.

The variation of a given parameter is not performed dynamically during the time evolution. Instead, discrete values within the chosen parameter interval are selected, and for each value the coupled equations of motion are solved independently using the same initial conditions described previously. After completing the numerical integration for one parameter value, the parameter is incremented by a small step and the equations are solved again. This procedure is repeated across the full parameter range of interest.

For each parameter value, the steady-state portions of the resulting time-domain solutions are Fourier transformed to obtain the corresponding frequency spectra. The collection of these spectra, assembled as a function of the scanned parameter, forms the contour plots presented in Figs.3-6. This approach ensures that the observed variations in comb structure arise solely from changes in the selected control parameter and not from transient or history-dependent effects.

\begin{figure}
    \centering
    \includegraphics[width=\linewidth]{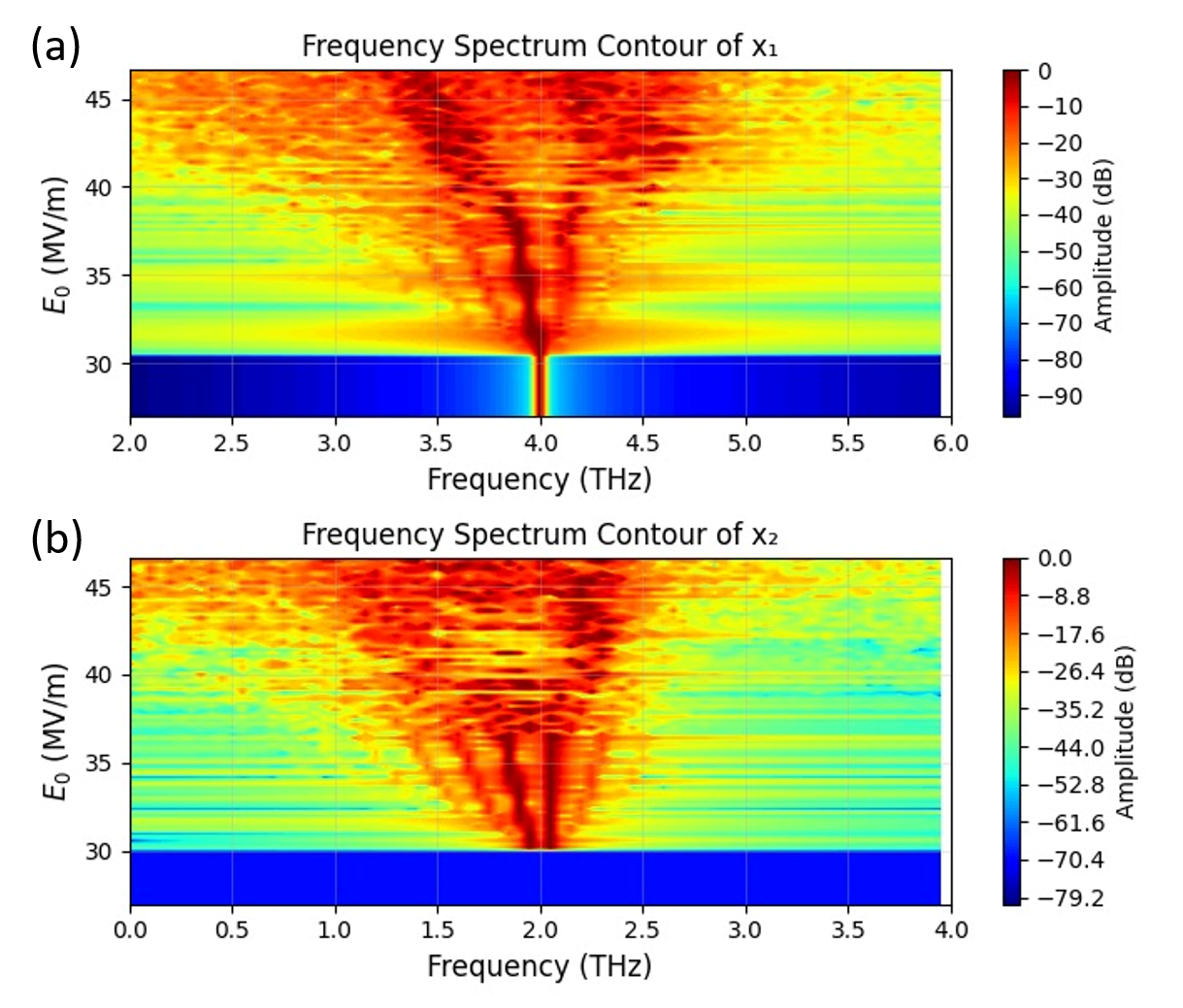}
    \caption{
    Contour plots showing the peak electric field amplitude $E_0$ dependence of the frequency-domain dynamics of the coupled Higgs-like and Goldstone-like phonon modes.
    Panel (a) shows the response of the Higgs-like mode $x_1$, while panel (b) shows the response of the Goldstone-like mode $x_2$.
    Only the peak electric field amplitude $E_0$ is varied within the range $\left[26.9507,\,46.6454\right]$,
    while all other parameters are fixed to their default values (see TABLE~II).}
    \label{fig1}
\end{figure}

\subsection{Electric Field Dependence}

Fig.~3 shows contour plots of the frequency-domain response as a function of the peak electric field amplitude $E_0$ for both phonon modes. The upper panel (a) corresponds to the Higgs-like mode ($x_1$), while the lower panel (b) shows the Goldstone-like mode ($x_2$). These plots illustrate how the phononic frequency combs evolve with increasing strength of the terahertz driving field.

For the Higgs-like mode, a clear threshold behavior is observed. Below an electric field strength of approximately $30~\mathrm{MV\,m^{-1}}$, the spectral response is dominated by a single peak near the bare Higgs-mode frequency of 4~THz, indicating nearly linear oscillations with negligible excitation of the Goldstone-like mode. In this regime, the nonlinear coupling is insufficient to transfer energy between the two phonon modes.

Once the electric field exceeds this threshold, additional frequency components appear symmetrically around the central peak, signaling the onset of phononic frequency comb formation. This threshold corresponds to the minimum driving strength required for the Higgs-like mode to reach amplitudes large enough to efficiently excite the Goldstone-like mode through nonlinear coupling. Consistently, the emergence of comb structures in the $x_2$ spectrum occurs only beyond the same threshold, confirming the coupled nature of the dynamics.

As the electric field strength is further increased, the number of comb lines grows and the spectral bandwidth broadens for both modes. The spacing between adjacent comb peaks also increases, reflecting enhanced nonlinear interactions at higher drive amplitudes. Beyond a certain electric field strength, however, the regular comb structure deteriorates. The evenly spaced peaks disappear and the spectrum becomes increasingly irregular, indicating the onset of strongly nonlinear or chaotic dynamics in which no well-defined phononic frequency comb is sustained.

\begin{figure}
    \centering
    \includegraphics[width=\linewidth]{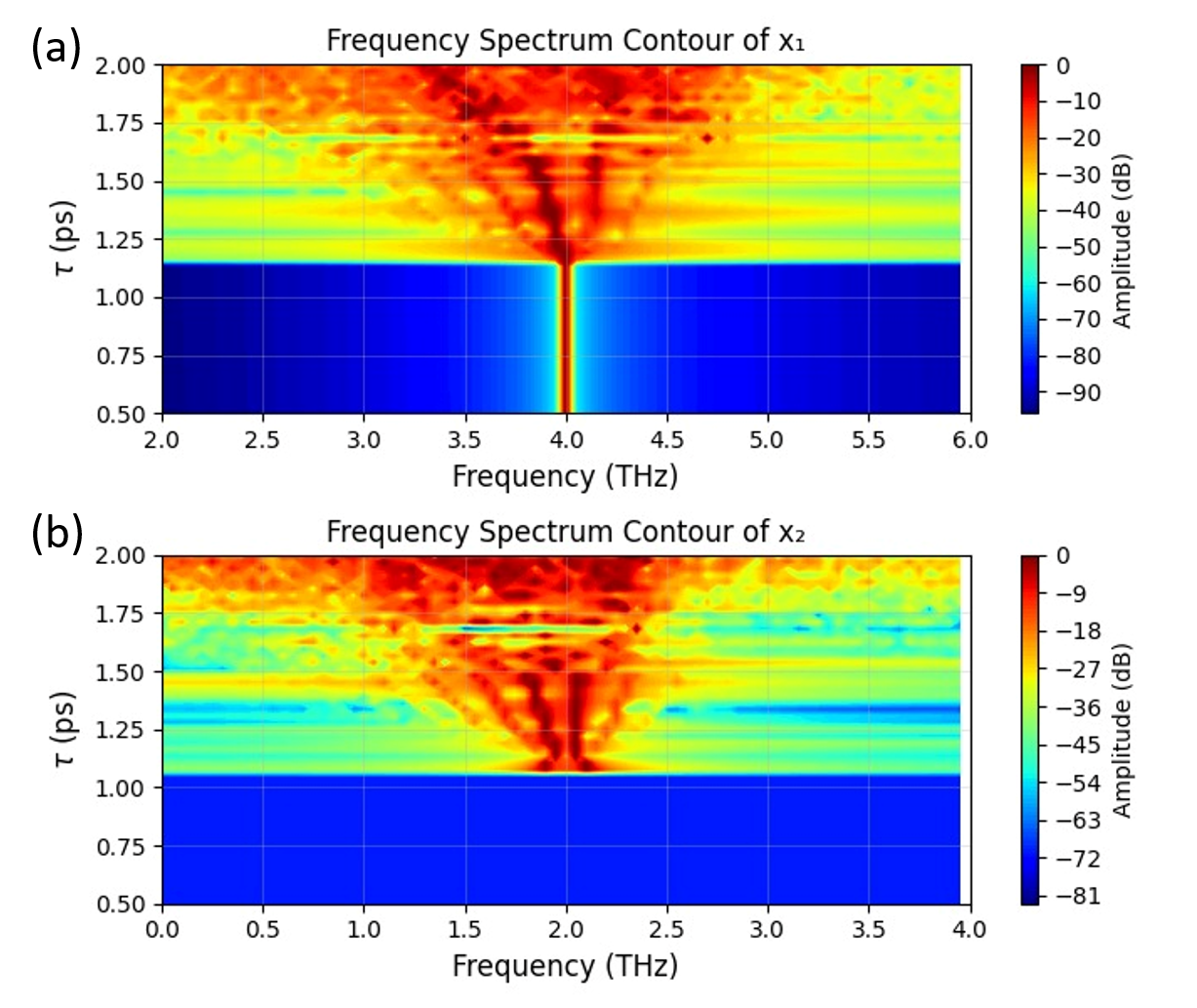}
    \caption{
    Contour plots showing the dependence of the frequency-domain dynamics of the coupled Higgs-like and Goldstone-like phonon modes on the pulse width $\tau$.
    Panel (a) shows the response of the Higgs-like mode $x_1$, while panel (b) shows the response of the Goldstone-like mode $x_2$.
    Only $\tau$ is varied within the range $[0.5,\,2.0]~\mathrm{ps}$,
    while all other parameters are fixed to their default values (see Table~II).}
    \label{fig1}
\end{figure}

\subsection{Pulse Width Dependence}

Fig.~4 shows contour plots of the frequency-domain response as a function of the pulse width $\tau$ for both phonon modes. These plots illustrate how the temporal duration of the terahertz driving pulse influences the formation and stability of phononic frequency combs.

For small values of $\tau$, the response of the Higgs-like mode is dominated by a single spectral peak near its bare frequency, while the Goldstone-like mode exhibits minimal or no excitation. In this regime, the driving pulse is too short-lived to transfer sufficient energy into the Higgs mode, thereby suppressing nonlinear energy transfer to the Goldstone-like mode.

As the pulse width exceeds a critical threshold value, additional frequency components appear in the spectrum of the Higgs-like mode, signaling the onset of phononic frequency comb formation. Concurrently, well-defined comb structures emerge in the Goldstone-like mode, indicating that the longer pulse duration enables the Higgs mode to reach amplitudes large enough to activate the nonlinear coupling between the two modes.

With further increases in $\tau$, the number of comb lines grows and the spectral bandwidth broadens for both modes. The spacing between adjacent peaks also increases, reflecting enhanced nonlinear modulation due to sustained driving of the lattice. Beyond an optimal pulse duration, however, the comb structure deteriorates. The spectral peaks become uneven and less distinct, and the response evolves toward irregular, noisy spectra characteristic of strongly nonlinear or chaotic dynamics.

These observations demonstrate that the pulse width serves as a crucial control parameter, governing both the onset of coherent phononic frequency combs and their eventual breakdown due to nonlinear instability.

\begin{figure}
    \centering
    \includegraphics[width=\linewidth]{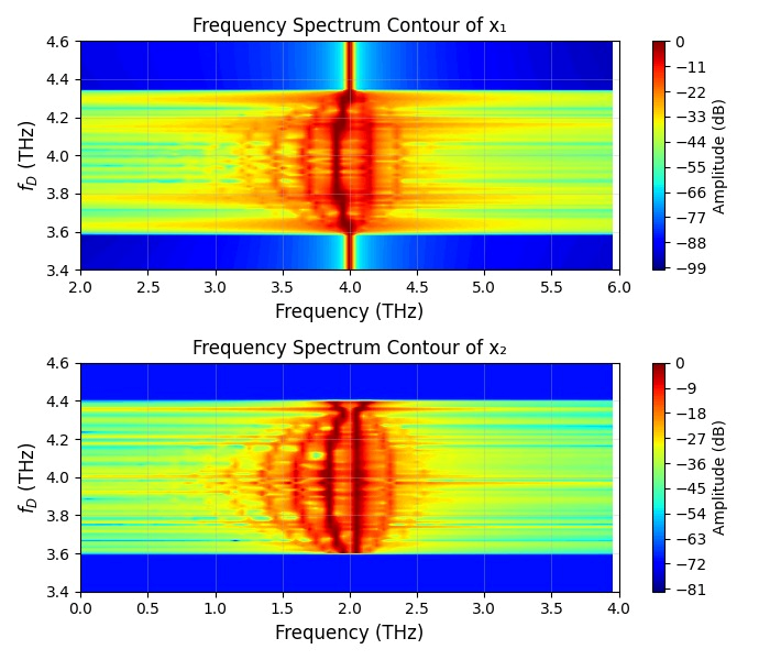}
    \caption{
    Contour plots showing the dependence of the frequency-domain dynamics of the coupled Higgs-like and Goldstone-like phonon modes on the driving frequency $f_D$.
    Panel (a) shows the response of the Higgs-like mode $x_1$, while panel (b) shows the response of the Goldstone-like mode $x_2$.
    Only $f_D$ is varied within the range $\left[3.4,\,4.6\right]~\mathrm{THz}$,
    while all other parameters are fixed to their default values (see Table~II).}
    \label{fig1}
\end{figure}
\subsection{Driving Frequency Dependence}

Fig.~5 presents contour plots of the frequency-domain response as a function of the driving frequency $f_D$ for both phonon modes. The results indicate that phononic frequency comb formation occurs only within a finite interval of driving frequencies, bounded by lower and upper thresholds. Outside this interval, the response is dominated by a single spectral peak at the bare frequency of the Higgs-like mode, corresponding to simple harmonic oscillations without comb generation.

Within the comb-forming frequency window, the most pronounced phononic frequency combs appear when the driving frequency is tuned close to the resonance of the Higgs-like mode, $f_D \approx 4~\mathrm{THz}$. At this optimal driving condition, the nonlinear coupling between the Higgs-like and Goldstone-like modes is strongest, resulting in the largest number of comb lines and the broadest spectral bandwidth. Both modes exhibit well-defined, evenly spaced spectral peaks, reflecting efficient nonlinear energy redistribution.

As the driving frequency is detuned away from resonance toward either threshold, the comb structure progressively weakens. The number of spectral peaks decreases, and the spacing between adjacent comb lines becomes smaller. Near the boundaries of the comb-forming region, the nonlinear response collapses to a single dominant peak, indicating that the parametric processes required for phononic frequency comb generation are no longer sustained.

These observations highlight the critical role of near-resonant driving in stabilizing coherent nonlinear phonon dynamics and underscore the importance of frequency tuning as a control parameter for phononic frequency comb generation.

\begin{figure}
    \centering
    \includegraphics[width=\linewidth]{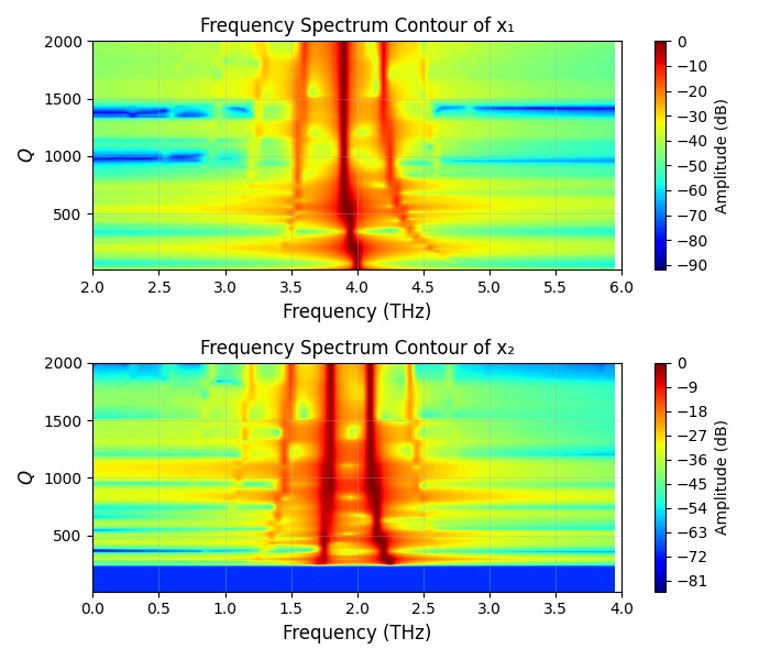}
    \caption{
    Contour plots showing the dependence of the frequency-domain dynamics of the coupled Higgs-like and Goldstone-like phonon modes on the quality factor $Q$.
    Panel (a) shows the response of the Higgs-like mode $x_1$, while panel (b) shows the response of the Goldstone-like mode $x_2$.
    The quality factors of both modes are taken to be equal ($Q_1 = Q_2 = Q$) and are varied within the range $[10,\,2000]$,
    while all other parameters are fixed to their default values (see Table~II).}
    \label{fig1}
\end{figure}

\subsection{Quality Factor Dependence}

Fig.~6 shows contour plots of the frequency-domain response as a function of the quality factor $Q$, which inversely characterizes the damping of both phonon modes. In this analysis, the quality factors of the Higgs-like mode ($x_1$) and the Goldstone-like mode ($x_2$) are taken to be equal and are varied simultaneously.

At low quality factor values, corresponding to strong damping, the system exhibits predominantly harmonic behavior. The Higgs-like mode oscillates near its bare frequency of approximately $4~\mathrm{THz}$, while the Goldstone-like mode remains weakly excited or effectively inactive. In this regime, dissipation suppresses the buildup of coherent nonlinear dynamics, preventing the formation of phononic frequency combs.

As the quality factor increases, damping is reduced and nonlinear effects become more pronounced. Initially, only a small number of spectral peaks appear, and the spacing between adjacent peaks is relatively large. With further increases in $Q$, additional comb lines emerge in both modes, and the spacing between peaks decreases. This behavior reflects the longer phonon lifetimes at higher quality factors, which allow sustained energy exchange between the coupled modes and enable higher-order nonlinear interactions.

Beyond a critical quality factor, the comb spacing becomes approximately constant, indicating the onset of a stable nonlinear regime. In this regime, increasing $Q$ primarily leads to an increase in the number of observable comb lines rather than significant changes in their spacing. Both the Higgs-like and Goldstone-like modes exhibit progressively richer comb structures, with the number of peaks increasing from a few discrete components at moderate $Q$ to multiple well-defined modes at high $Q$.

These observations highlight the essential role of low dissipation in enabling robust phononic frequency comb generation and underscore the importance of quality factor as a key control parameter for sustaining coherent nonlinear phonon dynamics.


\section{Conclusion}
In this work, we have studied phononic frequency comb generation in hexagonal InMnO$_3$ within a nonlinear phononics framework that captures the coupled dynamics of Higgs-like and Goldstone-like phonon modes. By numerically solving the nonlinear equations of motion under terahertz excitation, we demonstrated that coherent frequency combs arise from energy transfer mediated by nonlinear mode coupling. Our analysis shows that comb formation requires threshold values of both electric field strength and pulse duration, reflecting the need for sufficient excitation of the Higgs mode to activate the Goldstone mode. 

We further found that the comb structure is highly tunable: increasing drive strength or pulse width broadens the comb and increases the number of spectral lines, while the driving frequency must lie within a finite window around the Higgs resonance to sustain coherent comb formation. The role of damping is equally crucial, with higher quality factors enabling denser and more stable combs, whereas excessive driving or insufficient damping leads to spectral irregularity and loss of coherence. Overall, this study highlights the central role of nonlinear Higgs--Goldstone coupling in shaping phononic frequency combs and establishes a clear parameter space for their controlled generation in complex oxides, offering a pathway toward phonon-based frequency control in ultrafast solid-state systems.
\\
\bibliography{apssamp}

\end{document}